\newcommand \kms {{\rm km~s}^{-1}}

\newcommand \beqn {\begin{equation}}
\newcommand \eeqn {\end{equation}}

\documentclass[apj]{emulateapj}
\usepackage{epsfig}
\usepackage{natbib}

\begin{document}

\title{Comparison of Hectospec Virial Masses with SZE Measurements} 
\shorttitle{Hectospec Virial Masses and SZE}
\shortauthors{Rines, Geller, \& Diaferio}

\author{Kenneth Rines\altaffilmark{1,2}, Margaret J. Geller\altaffilmark{2}, 
and Antonaldo Diaferio\altaffilmark{3,4}} 
\email{kenneth.rines@wwu.edu}

\altaffiltext{1}{Department of Physics \& Astronomy, Western Washington University, Bellingham, WA 98225; kenneth.rines@wwu.edu}
\altaffiltext{2}{Smithsonian Astrophysical Observatory, 60 Garden St, Cambridge, MA 02138}
\altaffiltext{3}{Universit\`a degli Studi di Torino,
Dipartimento di Fisica Generale ``Amedeo Avogadro'', Torino, Italy}
\altaffiltext{4}{Istituto Nazionale di Fisica Nucleare (INFN), Sezione di Torino, Torino, Italy}

\begin{abstract}

We present the first comparison of virial masses of
galaxy clusters with their Sunyaev-Zel'dovich Effect (SZE) signals.  We study
15 clusters from the Hectospec Cluster Survey (HeCS) with
MMT/Hectospec spectroscopy and published SZE signals.  We measure
virial masses of these clusters from an average of 90 member redshifts
inside the radius $r_{100}$.  The virial masses of the clusters are
strongly correlated with their SZE signals (at the 99\% confidence
level using a Spearman rank-sum test).  This correlation suggests that
$Y_{SZ}$ can be used as a measure of virial mass.  Simulations predict
a powerlaw scaling of $Y_{SZ}\propto M_{200}^\alpha$ with
$\alpha\approx$1.6.  Observationally, we find $\alpha$=1.11$\pm$0.16,
significantly shallower (given the formal uncertainty) than the theoretical 
prediction.   However, the selection function of our sample is unknown 
and a bias against less massive clusters cannot be excluded (such a 
selection bias could artificially flatten the slope). 
Moreover, our sample indicates that  the relation between velocity 
dispersion (or virial mass estimate) and SZE signal has significant intrinsic 
scatter, comparable to the
range of our current sample. More detailed studies of scaling relations
are therefore needed to derive a robust determination of the relation between
cluster mass and SZE.

\end{abstract}


\section{Introduction}

Clusters of galaxies are the most massive virialized systems in the
universe.  The normalization and evolution of the cluster mass
function is therefore a sensitive probe of the growth of structure and
thus cosmology \citep[e.g.,][and references
therein]{cirsmf,rines08,vikhlinin09b,henry09,mantz08,rozo08}.  Many
methods exist to estimate cluster masses, including dynamical masses
from either galaxies \citep[][]{zwicky1937} or intracluster gas
\citep[e.g.,][]{flg80}, gravitational lensing \citep[e.g.,][]{smith05,richard10}, and the
Sunyaev-Zel'dovich effect \citep[SZE][]{sz72}.  In practice, these
estimates are often made using simple observables, such as velocity
dispersion for galaxy dynamics or X-ray temperature for the
intracluster gas.  If one of these observable properties of clusters
has a well-defined relation to the cluster mass, a large survey can
yield tight constraints on cosmological parameters
\citep[e.g.,][]{majumdar04}.  There is thus much interest in
identifying cluster observables that exhibit tight scaling relations
with mass \citep[][]{kravtsov06,rozo08}.  Numerical simulations
indicate that X-ray gas observables \citep{nagai07} and SZE signals
\citep{motl05} are both candidates for tight scaling
relations.  Both methods are beginning to gain observational support
\citep[e.g.,][]{henry09,lopes09b,mantz09b,locutushuang09}.  Dynamical 
masses from galaxy velocities are unbiased in numerical simulations 
\citep{diaferio1999,evrard07}, and recent results from hydrodynamical 
simulations indicate that virial masses may have scatter as small as 
$\sim$5\% \citep{lau10}.  

Previous studies have compared SZE signals to hydrostatic X-ray masses \citep{bonamente08,plagge10} and gravitational lensing masses \citep[][hereafter M09]{marrone09}.
Here, we make the first comparison between virial masses of galaxy
clusters and their SZE signals.  We use SZE measurements from the
literature and newly-measured virial masses of 15 clusters from
extensive MMT/Hectospec spectroscopy.  This comparison tests the
robustness of the SZE as a proxy for cluster mass and the physical
relationship between the SZE signal and cluster mass.  Large SZ cluster surveys are underway and are beginning to yield cosmological constraints \citep{carlstrom10,hincks10,staniszewski09}.

We assume a cosmology of
$\Omega_m$=0.3, $\Omega_\Lambda$=0.7, and $H_0$=70 km s$^{-1}$ Mpc$^{-1}$ for all calculations.

\section{Observations}

\subsection{Optical Photometry and Spectroscopy}

We are completing the Hectospec Cluster Survey (HeCS), a study of an
X-ray flux-limited sample of 53 galaxy clusters at moderate redshift
with extensive spectroscopy from MMT/Hectospec.  HeCS includes all
clusters with ROSAT X-ray  fluxes of $f_X>5\times10^{-12}$erg s$^{-1}$ at [0.5-2.0]keV from the Bright Cluster Survey \citep[BCS][]{bcs} or REFLEX survey \citep{reflex} with optical imaging 
in the Sixth Data Release (DR6) of SDSS \citep{dr6}.
We use DR6 photometry to select Hectospec targets.
The HeCS targets are all brighter than $r$=20.8 (SDSS catalogs are 95\%
complete for point sources to $r$$\approx$22.2).
Out of the HeCS sample, 15 clusters have published SZ measurements.

\subsubsection{Spectroscopy: MMT/Hectospec and SDSS}

HeCS is a spectroscopic survey of clusters in the redshift range
0.10$\leq$$z$$\leq$0.30.  We measure spectra with the
Hectospec instrument \citep{hectospec} on the MMT 6.5m telescope.
Hectospec provides simultaneous spectroscopy of up to 300 objects
across a diameter of 1$^\circ$.  This telescope and instrument
combination is ideal for studying the virial regions and outskirts of
clusters at these redshifts.  We use the red sequence to preselect likely
cluster members as primary targets, and we fill fibers with bluer
targets (Rines et al.~in prep.~describes the details of target
selection).  
We eliminate all targets with existing SDSS spectroscopy from our target 
lists but include these in our final redshift catalogs.

\begin{figure*} 
\figurenum{1} 
\plotone{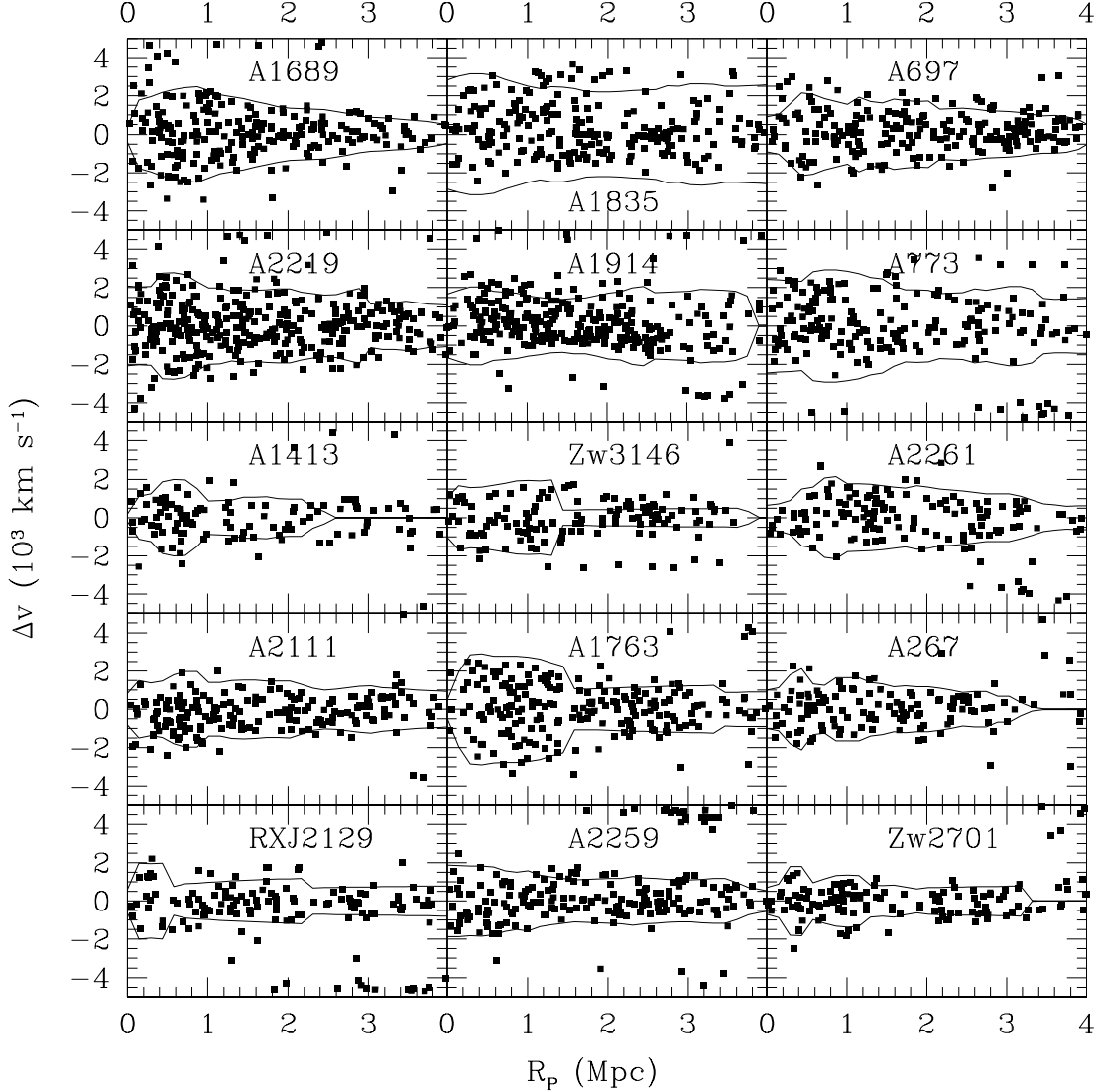}
\caption{\label{hecsyszcaus} Redshift versus projected 
clustrocentric radius for the 15 HeCS clusters studied here.  Clusters are ordered left-to-right and top-to-bottom by decreasing values of $Y_{SZ}D_A^2(r_{2500})$.  The solid lines show the locations of the caustics,
which we use to identify cluster members.  The Hectospec data extend
out to $\sim$8 Mpc; the figure shows only the inner
4 Mpc to focus on the virial regions.  }
\end{figure*}

Of the 15 clusters studied here, one was observed with a single
Hectospec pointing and the remaining 14 were observed with two
pointings.  Using multiple pointings and incorporating SDSS redshifts
of brighter objects mitigate fiber collision issues.  Because the
galaxy targets are relatively bright ($r$$\leq$20.8), the spectra were
obtained with relatively short exposure times of 3x600s to 4x900s
under a variety of observing conditions.

Figure \ref{hecsyszcaus} shows the redshifts of galaxies versus their
projected clustrocentric radii for the 15 clusters studied here.
The infall patterns are clearly present in all clusters.  We use the
caustic technique
\citep{diaferio1999} to determine cluster membership.  Briefly, the
caustic technique uses a redshift-radius diagram to isolate cluster
members in phase space by using an adaptive kernel estimator to smooth
out the galaxies in phase space, and then determining the edges of
this distribution \citep[see][for a recent review]{diaferio09}.  This
technique has been successfully applied to optical studies of X-ray
clusters, and yields cluster mass estimates in agreement with
estimates from X-ray observations and gravitational lensing
\citep[e.g.,][and references therein]{cairnsi,bg03,diaferio05,cirsi,cirsmf}.

We apply the prescription of \citet{danese} to determine the mean
redshift $cz_\odot$ and projected velocity dispersion $\sigma_p$ of
each cluster from all galaxies within the caustics.  We calculate
$\sigma_p$ using only the cluster members projected within $r_{100}$
estimated from the caustic mass profile.  

\subsection{SZE Measurements}

The SZE detections are primarily from \citet[][hereafter
B08]{bonamente08}, supplemented by three measurements from
\citet[][hereafter M09]{marrone09}.  Most of the SZ data were obtained
with the OVRO/BIMA arrays; the additional clusters from M09 were
observed with the Sunyaev-Zel'dovich Array \citep[SZA;
e.g.,][]{muchovej07}.

Numerical simulations indicate that the integrated Compton y-parameter
$Y_{SZ}$ has smaller scatter than the peak y-decrement $y_{peak}$
\citep{motl05}, so B08 and M09 report only $Y_{SZ}$. Although
$y_{peak}$ should be nearly independent of redshift, $Y_{SZ}$ depends
on the angular size of the cluster.  The quantity $Y_{SZ}D_A^2$
removes this dependence.  Thus, we compare our dynamical mass
estimates to this quantity rather than $y_{peak}$ or $Y_{SZ}$.  Table
\ref{hecsysztab} summarizes the SZ data and optical spectroscopy.

It is also critical to determine the radius within which
$Y_{SZ}$ is determined.  B08 use $r_{2500}$, the radius
that encloses an average density of 2500 times the critical density at
the cluster's redshift; $r_{2500}$ has physical values of 300-700
kpc for the massive clusters studied by
B08 (470-670 kpc for the subsample studied
here).  M09 use a physical radius of 350 kpc
because this radius best matches their lensing data.  

To use both sets of data, we must estimate the conversion between
$Y_{SZ}(r_{2500})$ measured within $r_{2500}$ and
$Y_{SZ}(r=350~\mbox{kpc})$ measured within the smaller radius
$r=$350~kpc.  There are 8 clusters analyzed in both B08 and M09 (5 of
which are in HeCS).  We perform a least-squares fit to
$Y_{SZ}(r_{2500})-Y_{SZ}(r=350\mbox{kpc})$ to determine an approximate
aperture correction for the M09 clusters.  We list both quantities in
Table 1.

\begin{table*}[th] \footnotesize
\begin{center}
\caption{\label{hecsysztab} \sc HeCS Dynamical Masses and SZE Signals}
\begin{tabular}{lcccccccc}
\tableline
\tableline
\tablewidth{0pt}
Cluster & $z$    & $\sigma_p$ & $M_{100,v}$ & $M_{100,c}$ & $Y_{SZ}D_A^2$ & $Y_{SZ}D_A^2$ & SZE \\ 
 &    &     &                   &                   &   (350 kpc) & $(r_{2500})$ & \\
 &    & $\kms$ & $10^{14} M_\odot$ & $10^{14} M_\odot$ & $10^{-5}$Mpc$^{-2}$ & $10^{-4}$Mpc$^2$ &  Ref.\\ 
\tableline
          A267 &  0.2288 &  $ 743 ^{+81} _{-61}$  & 6.86$\pm$0.82 & 4.26$\pm$0.14 & 3.08$\pm$0.34 & 0.42$\pm$0.06 & 1  \\ 
          A697 &  0.2812 &   $ 784 ^{+77} _{-59}$  & 6.11$\pm$0.69 & 5.96$\pm$3.51 & -- & 1.29$\pm$0.15  & 1  \\ 
          A773 &  0.2174 &   $ 1066 ^{+77} _{-63}$ & 18.4$\pm$1.7  & 16.3$\pm$0.7 & 5.40$\pm$0.57 & 0.90$\pm$0.10 & 1  \\ 
        Zw2701 &  0.2160 &   $ 564 ^{+63} _{-47}$  & 3.47$\pm$0.42 & 2.69$\pm$0.30 & 1.46$\pm$0.016 & {0.17$\pm$0.02}$^\tablenotemark{a}$ & 2  \\ 
        Zw3146 &  0.2895 &   $ 752 ^{+92} _{-67}$  & 6.87$\pm$0.89 & 4.96$\pm$0.91 & -- & 0.71$\pm$0.09 & 1  \\ 
         A1413 &  0.1419 &  $ 674 ^{+81} _{-60}$  & 6.60$\pm$0.85 & 3.49$\pm$0.15 & 3.47$\pm$0.24 & 0.81$\pm$0.12 & 1  \\ 
         A1689 &  0.1844 &   $ 886 ^{+63} _{-52}$  & 15.3$\pm$1.4 & 9.44$\pm$5.66 & 7.51$\pm$0.60 & 1.50$\pm$0.14 & 1  \\ 
        A1763  &  0.2315 &   $ 1042 ^{+79} _{-64}$  & 16.9$\pm$1.6 & 12.6$\pm$1.5 & 3.10$\pm$0.32 & {0.46$\pm$0.05}$^\tablenotemark{a}$ & 2  \\ 
         A1835 &  0.2507 &   $ 1046 ^{+66} _{-55}$ & 19.6$\pm$1.6 & 20.6$\pm$0.3 & 6.82$\pm$0.48 & 1.37$\pm$0.11 & 1  \\ 
         A1914 &  0.1659 &   $ 698 ^{+46} _{-38}$  & 6.70$\pm$0.57 & 6.21$\pm$0.21 & -- & 1.08$\pm$0.09 & 1  \\ 
         A2111 &  0.2290 &   $ 661 ^{+57} _{-45}$  & 4.01$\pm$0.41 & 4.77$\pm$1.23 & -- & 0.55$\pm$0.12 & 1  \\ 
         A2219 &  0.2256 &   $ 915 ^{+53} _{-45}$  & 12.8$\pm$1.0 & 12.0$\pm$4.7 & 6.27$\pm$0.26  & {1.19$\pm$0.05}$^\tablenotemark{a}$  & 2  \\ 
         A2259 &  0.1606 &  $ 735 ^{+67} _{-53}$  & 5.59$\pm$0.60 & 4.90$\pm$1.69 & --  & 0.27$\pm$0.10 & 1  \\ 
         A2261 &  0.2249 &  $ 725 ^{+75} _{-57}$  & 7.13$\pm$0.83 & 5.10$\pm$2.07 & -- & 0.71$\pm$0.09 & 1  \\ 
       RXJ2129 &  0.2338 &  $ 684 ^{+88} _{-64}$  & 4.31$\pm$0.57 & 2.94$\pm$0.13 & -- & 0.40$\pm$0.07 & 1  \\ 
\tableline
\end{tabular}
\end{center}
\tablecomments{$^a$Extrapolated to $r_{2500}$ using the best-fit relation between $Y_{SZ}D_A^2(350\mbox{kpc})$ and $Y_{SZ}D_A^2(r_{2500})$
 for eight clusters in common between B08 and M09.}
\tablecomments{Redshift $z$ and velocity dispersion $\sigma_p$ are computed for galaxies defined as members using the caustics.  Masses $M_{100,v}$ and $M_{100,c}$ are evaluated using the virial mass profile and caustic mass profile respectively. }
\tablecomments{REFERENCES: SZE data are from (1) Bonamente et al.~2008 and (2) Marrone et al.~2009.  
}
\end{table*}

\section{Results}

\begin{figure*} 
\figurenum{2} 
\plotone{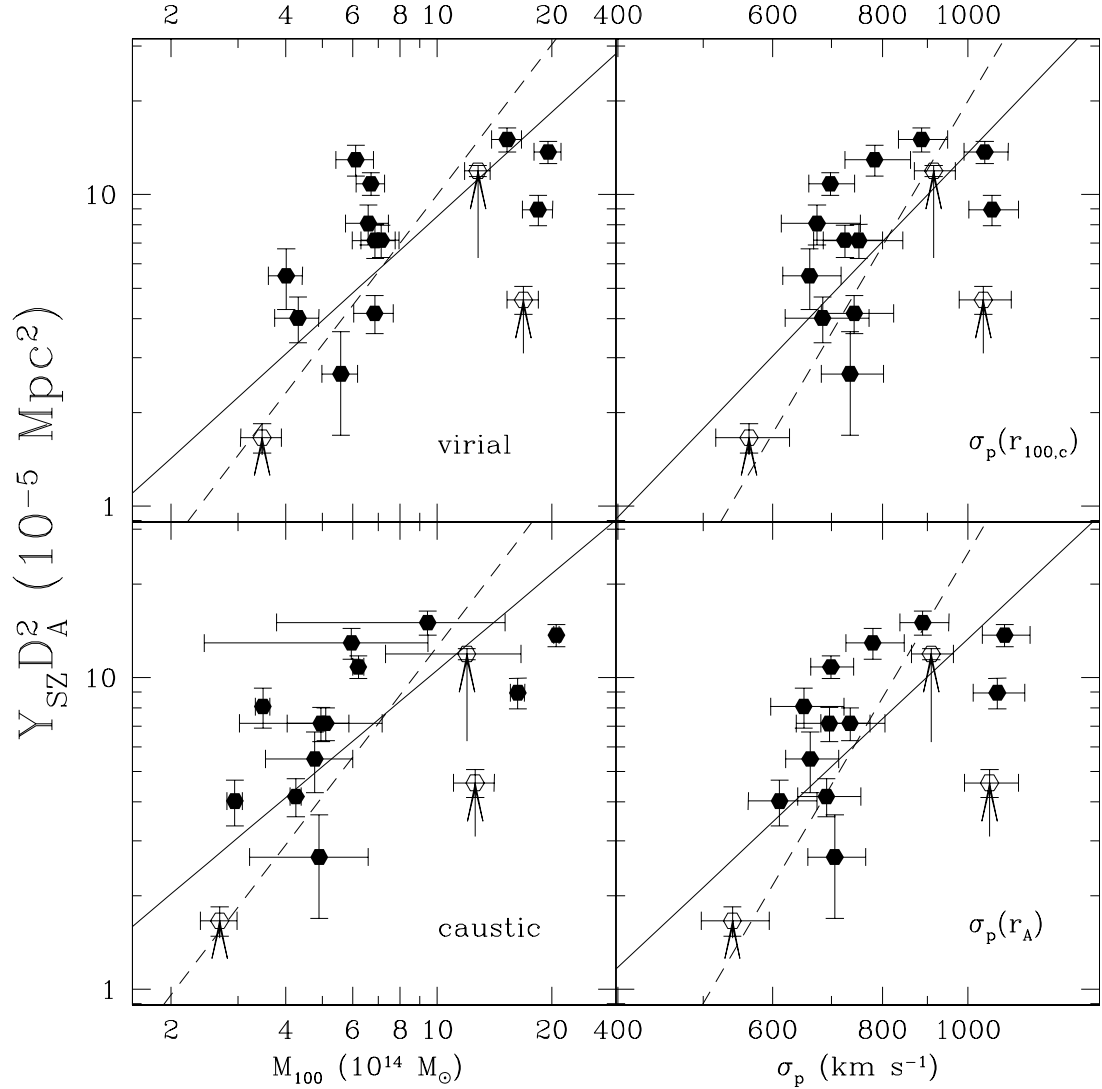}
\caption{\label{hecsysz} Integrated S-Z Compton parameter $Y_{SZ}D_A^2$
versus dynamical properties for 15 clusters from HeCS. {\it Left panels:} 
SZE data versus virial mass $M_{100}$ estimated from the virial mass profile 
(top) and the caustic mass profile (bottom).  Solid and open points 
indicate SZ measurements from B08 and M09
respectively.  The dashed line shows the slope of the scaling predicted 
from numerical simulations: $Y_{SZ}\propto M^{1.6}$ \citep{motl05}, 
while the solid line shows the ordinary least-squares bisector.  Arrows 
show the aperture corrections to the SZE measurements (see text).
{\it Right panels:} SZE data versus projected velocity dispersions 
measured for galaxies inside the caustics and (top) inside $r_{100,c}$ 
estimated from the caustic mass profile and (bottom) inside the Abell 
radius 2.14 Mpc.  The dashed line shows the scaling predicted 
from simulations: $Y_{SZ}\propto M^{1.6}$ \citep{motl05} and $\sigma 
\propto M^{0.33}$ \citep{evrard07}.  The solid line shows the ordinary 
least-squares bisector.  Data points and arrows are defined as in the left panels. 
} 
\end{figure*}

We examine two issues: (1) the strength of the correlation between SZE
signal and the dynamical mass and (2) the slope of the relationship
between them.  Figure \ref{hecsysz} shows the $Y_{SZ}-\sigma_p$
relation.  Here, we compute $\sigma_p$ for all galaxies inside both
the caustics and the radius $r_{100,c}$ defined by the caustic mass
profile [$r_\delta$ is the radius within which the enclosed density is
$\delta$ times the critical density $\rho_c(z)$].  

Because we make the first comparison of dynamical properties and SZE
signals, we first confirm that these two
variables are well correlated.  A nonparametric Spearman rank-sum test
(one-tailed) rejects the hypothesis of uncorrelated data at the 98.4\%
confidence level.  The strong correlation in the data suggests that
both $\sigma_p$ and $Y_{SZ}D_A^2$ increase with increasing cluster
mass.

Hydrodynamic numerical simulations indicate that $Y_{SZ}$ (integrated
to $r_{500}$) scales with cluster mass as $Y_{SZ}\propto
M_{500}^\alpha$, where $\alpha$=1.60 with radiative cooling and star
formation, and 1.61 for simulations with radiative cooling, star
formation, and AGN feedback \citep[$\alpha$=1.70 for non-radiative
simulations,][]{motl05}.  Combining this result with the virial
scaling relation of dark matter particles, $\sigma_p\propto
M_{200}^{0.336\pm0.003}$ \citep{evrard07}, the expected scaling is
$Y_{SZ}\propto \sigma^{4.76}$ (we assume that $M_{100}\propto
M_{500}$).  The right panels of Figure \ref{hecsysz} shows this
predicted slope (dashed lines).

The bisector of the least-squares fits to the data has a slope of 
$2.94\pm0.74$, significantly shallower than the predicted slope of 4.8.  

We recompute the velocity dispersions $\sigma_{p,A}$ for all galaxies
within one Abell radius (2.14 Mpc) and inside the
caustics.  Surprisingly, the correlation is slightly stronger (99.4\%
confidence level).  This result supports the idea that velocity
dispersions computed within a fixed physical radius retain strong
correlations with other cluster observables, even though we measure
the velocity dispersion inside different fractions of the virial
radius for clusters of different masses.  Because cluster velocity
dispersions decline with radius
\citep[e.g.][]{cairnsi,cirsi}, $\sigma_{p,A}$ may be smaller than
$\sigma_{p,100}$ (measured within $r_{100,c}$) for low-mass clusters,
perhaps exaggerating the difference in measured velocity dispersions
relative to the differences in virial mass (i.e., $\sigma_{p,A}$ of a
low-mass cluster may be measured within 2$r_{100}$ while
$\sigma_{p,A}$ of a high-mass cluster may be measured within
$r_{100}$; the ratio $\sigma_{p,A}$ of these clusters would be
exaggerated relative to the ratio $\sigma_{p,100}$).  Future cluster
surveys with enough redshifts to estimate velocity dispersions but too
few to perform a caustic analysis should still be sufficient for
analyzing scaling relations.

Because of random errors in the mass estimation, the virial mass and
the caustic mass within a given radius do not necessarily coincide.
Therefore, the radius $r_{100}$ depends on the mass estimator used.
Figure \ref{hecsysz} shows the scaling relations for two estimated
masses $M_{100,c}$ and $M_{100,v}$; $M_{100,c}$ is the mass estimated
within $r_{100,c}$ (where both quantities are defined from the caustic
mass profile), and $M_{100,v}$ is the mass estimated within
$r_{100,v}$ \citep[both quantities are estimated with the virial
theorem, e.g.,][]{cirsi}.
including galaxies projected inside $r_{100,v}$.  
Similar to $\sigma_p$, there is a clear correlation between
$M_{100,v}$ and $Y_{SZ}D_A^2$ (99.0\% confidence with a Spearman
test).  The strong correlation of dynamical mass with SZE also holds for
$M_{100,c}$ estimated directly from the caustic technique (99.8\%
confidence).  

The bisector of the least-squares fits has a slope of $1.11\pm0.16$,
again significantly shallower than the predicted slope of 1.6.  
This discrepancy has two distinct origins.  By looking at the distribution
of the SZE signals in Figure \ref{hecsysz},  we see that, at a given 
velocity dispersion or mass, the SZE signals have a scatter which is a 
factor of $\sim$2.  Alternatively, at fixed SZE signal, there is a scatter of 
a factor of $\sim$2 in estimated virial mass. Unless
the observational uncertainties are significantly underestimated,
the data show substantial intrinsic scatter.
Moreover, this scatter is comparable to the range
of our sample  and, therefore, the error on the slope derived from our least-squares fit to
the data is likely to be underestimated \citep[see][for a detailed discussion of a Bayesian approach to fitting relations with measurement uncertainties and intrinsic scatter in both quantities]{andreon10}. 

Our shallow slopes may also arise in part from the fact that our
sample, which has been assembled from the literature and whose
selection function is difficult to determine, is likely
to be biased against clusters with small mass and low
SZE signal.   
Larger samples should determine
whether unknown observational biases or issues in the physical
understanding of the relation account for this discrepancy.

\section{Discussion}


The strong correlation between masses from galaxy dynamics and SZE
signals indicates that the SZE is a reasonable proxy for cluster mass.
B08 compare SZE signals to X-ray observables, in particular the
temperature $T_X$ of the intracluster medium and $Y_X=M_{gas}T_X$,
where $M_{gas}$ is the mass of the ICM \citep[see also][]{plagge10}.  
Both of these quantities are
measured within $r_{500}$, a significantly smaller radius than
$r_{100}$ where we measure virial mass.  M09 compare SZE signals to
masses estimated from gravitational lensing measurements.  The lensing
masses are measured within a radius of 350 kpc.  For the
clusters studied here, this radius is smaller than $r_{2500}$ and much
smaller than $r_{100}$.  Numerical simulations indicate that the
scatter in masses measured within an overdensity $\delta$ decreases as
$\delta$ decreases \citep[][]{white02}, largely because variations in
cluster cores are averaged out at larger radii.  Thus, the dynamical
measurement reaching to larger radius may provide a more robust
indication of the relationship between the SZE measurements and
cluster mass.  

The $Y_{SZ}D_A^2-M_{lens}$ data presented in M09 show a 
weaker correlation than our optical dynamical properties.  A Spearman 
test rejects the hypothesis of uncorrelated data for the M09 data at 
only the 94.8\% confidence level, compared to the 98.4-99.8\% confidence 
levels for our optical dynamical properties.  One possibility
is that $M_{lens}$ is more strongly affected by substructure in cluster 
cores and by line-of-sight structures than are the virial masses and 
velocity dispersions we derive. 

Few measurements of SZE at large radii ($>r_{500}$) are currently available.  
Hopefully, future SZ data will allow a comparison between virial mass and 
$Y_{SZ}$ within similar apertures.  

\section{Conclusions}

Our first direct comparison of virial masses, velocity dispersions,
and SZ measurements for a sizable cluster sample demonstrates a
strong correlation between these observables (98.4-99.8\% confidence).
The SZE signal increases with cluster mass.  However, the slopes of
both the $Y_{SZ}-\sigma$ relation ($Y_{SZ}\propto
\sigma_p^{2.94\pm0.74}$) and the $Y_{SZ}-M_{100}$ relation
($Y_{SZ}\propto M_{100}^{1.11\pm0.16}$) are significantly shallower
(given the formal uncertainties) than the slopes predicted by numerical 
simulations (4.76 and 1.60 respectively).  

This result may be partly explained by a bias against less massive clusters 
that could artificially flatten our measured slopes.
Unfortunately, the selection function of our sample is unknown and we are unable
to quantify the size of this effect. More importantly, our sample indicates that
the relation between SZE and virial mass estimates (or velocity dispersion) has a non-negligible intrinsic scatter.
A complete, representative cluster sample is required to robustly determine the 
size of this scatter, its origin, and its possible effect on the SZE as a mass proxy.

Curiously, $Y_{SZ}$ is more strongly correlated with both $\sigma_p$ and
$M_{100}$ than with $M_{lens}$ (M09).  Comparison of lensing masses
and cluster velocity dispersions (and virial masses) for larger,
complete, objectively selected samples of clusters may resolve these
differences. 

The full HeCS sample of 53 clusters will provide a large sample of
clusters with robustly measured velocity dispersions and virial masses
as a partial foundation for these comparisons.

\acknowledgements

We thank Stefano Andreon for fruitful discussions about fitting scaling 
relations with measurement errors and intrinsic scatter in both quantities.  
AD gratefully acknowledges partial support from INFN grant PD51.  We 
thank Susan Tokarz for reducing the spectroscopic data and Perry 
Berlind and Mike Calkins for assisting with the observations.


{\it Facilities:} \facility{MMT (Hectospec)}


\end{document}